\documentclass[aps,twocolumn,a4paper,showpacs]{revtex4}
\usepackage[colorlinks=true, pdfstartview=FitV, linkcolor=blue, citecolor=red, urlcolor=magenta,breaklinks=true]{hyperref}
\usepackage[T1,T5]{fontenc}
\usepackage{graphicx}
\usepackage[all]{xy}
\usepackage{amsmath}
\usepackage{amssymb}
\usepackage{color}
\usepackage{subeqnarray}
\usepackage{subfigure}
\usepackage{epstopdf}
\newcommand{\pa}{\partial}

\newcommand{\bes}{\begin{subequations}}
\newcommand{\ees}{\end{subequations}}
\def\ben{\begin{eqnarray}}
\def\een{\end{eqnarray}}
\newcommand{\bens}{\begin{subeqnarray}}
\newcommand{\eens}{\end{subeqnarray}}

\def\be{\begin{equation}}
\def\ee{\end{equation}}
\def\As{A\!\!\!/}

\def\ks{k\!\!\!/}

\def\ps{p\!\!\!/}

\def\ds{\partial\!\!\!/}

\begin{document}
\title{Podolsky electrodynamics from a condensation of topological defects}  
\author{Diego R. Granado$^{b}$}
\author{Antonio J. G. Carvalho$^{c}$}
\author{A. Yu. Petrov$^{c}$}
\author{Paulo J. Porfirio$^{a,c}$}
\affiliation{$^{a}$Department of Physics and Astronomy, University of Pennsylvania, Philadelphia, PA 19104, USA}
\affiliation{$^{b}$ Institute of Research and Development, Duy T{a}n University, 3 Quang Trung, H{a}i Ch{a}u, D{a} Nang, Vietnam}
\affiliation{$^{c}$ Departamento de F\'{i}sica, Universidade Federal da Para\'{i}ba, Caixa Postal 5008, 58051-970, Jo\~{a}o Pessoa, Para\'{i}ba, Brazil}

\pacs{11.27.+d, 11.10.Lm,  03.50.Kk}
%\date{\today}

\begin{abstract}
In this paper we demonstrate the arising of higher-derivative contributions to the effective action of electrodynamics on the base of generalized Julia-Toulouse mechanism and explicitly show that the complete effective action {generated} within this methodology is nonlocal.
\end{abstract}
\maketitle

%%%%%%%%%%%%%%%%%%%%%%%

%%%%%%%%%%%%%%%%%%%%%%%%%%%%%%%%%%%
\section{Introduction}

The nonlocality is treated now as an important ingredient in field theory models. Being initially introduced in order to take into account finite-size effects in phenomenology allowing to rule out ultraviolet divergences \cite{Efimov}, it began recently to acquire strong interest within other contexts, especially, within gravity, where there is a strong hope that namely nonlocal extension could allow to construct a gravity model both renormalizable (or even ultraviolet finite) and ghost-free, see discussion in \cite{Modesto}. The key idea of nonlocal field theories looks like follows. While the quadratic action of usual higher-derivative theories is described by a polynomial function of d'Alembertian operator which can be expanded in primitive multipliers, {and it} allows in arising of new, ghost degrees of freedom \cite{Hawking}, one can consider a class of theories where, instead of the polynomial function of $\Box$, the quadratic action is characterized by an essentially non-polynomial, so-called entire function of $\Box$, for example, the exponential one, which does not admit expansion in primitive multipliers and hence does not generate new {degrees} of freedom, see \cite{Modesto} and references therein. Various aspects of nonlocal field theories have been studied, including the exact solutions within the gravitational (mostly cosmological) context (see f.e. \cite{Biswas}) and explicit calculations of loop corrections in usual and supersymmetric theories \cite{quantNL}. Therefore, it is natural to expect that nonlocality can imply interesting physical effects within other contexts as well.

In this letter, we propose a generalization of the Julia-Toulouse (JT) mechanism to a nonlocal case. Indeed, it is well known \cite{CW} that the Julia-Toulouse mechanism is based on the coupling of the gauge field to some extra fields (topological defects) represented by some sources $J^{\mu}$. The idea behind {the} JT mechanism is that a proliferation of the topological defects in a system, such that they become dynamical fields (i.e. their condensation), drives the {system} to a phase transition. As it will be formally presented in the next section, superconductors can be seen as a good example to {illustrate} this idea. In the case of a superconductor, a proliferation of defects (vortices) can drive the system from a superconductivity state to a free Maxwell theory. In order to achieve this result we introduce a so-called activation term $J^{\mu}{\cal O}J_{\mu}$, and the integration over the topological defects implies in the arising of new terms modifying the {original} gauge theory. Initially, in \cite{CW} and further in \cite{CWours}, the operator ${\cal O}$ was suggested to {describe} only the low- energy fluctuations, that is, being proportional to an unit operator, thus giving a Thirring-like interaction. In this paper, we take into account not only low energy fluctuation but we also consider the UV fluctuations and show that as a {consequence,} the condensation of topological defects drives the Maxwell electrodynamics to a non-local Podolsky electrodynamics. 

This letter is organized as follows: in the section \ref{secI}, we, basing on \cite{Braga:2016atp}, provide a general description of the condensation of topological defects in regular superconductors; 
in the section \ref{emergentnonlocalterm}, we show how the mechanism used to describe superconductors can be used to obtain non-local Podolsky electrodynamics; in the section \ref{sec-cs}, we show how the non-local JT mechanism can also be applied in the Chern-Simons theory to generate a higher-derivative CS theory; finally, in the section \ref{sec-com} we present our comments and conclusions.

%==============================================================================
\section{Phase transition and the condensation of topological defects}
\label{secI}
%==============================================================================
In this section we {describe} the mechanism presented in \cite{Braga:2016atp} to explain how a condensation of topological defects can drive a system to a phase transition.
{Let us} consider an Euclidean action describing the electromagnetic field interacting with an external source with electric charge $q$
\begin{align}
\label{maxwell}
S_{em} = \int d^4x \left( \frac{1}{4} F_{\mu\nu}F_{\mu\nu} - iqA_{\mu}J_{\mu}  \right)
\end{align}
where $F_{\mu\nu} = \partial_{\mu}A_{\nu} - \partial_{\nu}A_{\mu}$ and $J_{\mu}$ is a classical current. The partition function of the system coupled with an external auxiliary current $j_{\mu}$, carrying charge $e$, summed over the classical configurations of the external sources $J_{\mu}$ reads
\begin{widetext}
 \begin{eqnarray}
 \label{pathint-maxwell}
Z[j]&=&\sum_{J}  \int {\cal D} A \delta[\partial_{\mu}J^{\mu}] e^{-\int d^4x \left( \frac{1}{4} F_{\mu\nu}F_{\mu\nu} - iqA_{\mu}J_{\mu}\right)} e^{-ie\int d^4x \; A_{\mu}j_{\mu}  }\nonumber\\
 &=&  \sum_{J}  \int {\cal D} A {\cal D} \theta e^{-\int d^4x \left( \frac{1}{4} F_{\mu\nu}F_{\mu\nu} - iq \left( A_{\mu} + \frac{1}{q} \partial_{\mu}\theta \right) J_{\mu}\right)} e^{-ie\int d^4x \; A_{\mu}j_{\mu}  }
\label{pathint-maxwell-2}
 \end{eqnarray}
\end{widetext}
In order to maintain {explicitly} the gauge invariance we inserted a delta function {requiring} the classical source to be conserved and exponentiated it with the help of an auxiliary field $\theta$. The gauge symmetry is realized as
\begin{align}
 \label{gaugesymmetry}
A_{\mu} &\rightarrow  A_{\mu}  + \partial_{\mu} \chi \nonumber\\
\theta &\rightarrow \theta - q\chi
 \end{align}
As a consequence of gauge invariance (current conservation) we have that 
\begin{equation}
J^\mu(x)=\int d\tau\frac{d y^\mu}{d\tau}\delta^4(x-y(\tau))
\label{delta}
\end{equation}
The summation over $J$ in \eqref{pathint-maxwell-2} represents the sum over all the possible worldlines $y(\tau)$ of charges. 
In \eqref{pathint-maxwell-2} we did not consider {dynamics of charges yet}, but soon we will supplement the action  with {the corresponding term given by $S(y(\tau))$}. For many point charges we have a sum over the worldlines of all the charges. For a continuous distribution of charges we would have a continuous source, whose sum over different ensemble configurations is defined by a path integral weighted by an action $S(J)$. The condensation is an operation that maps an ensemble of 1-currents into an ensemble of 1-forms. This operation specifies a physical process that connects different theories, describing the system in different phases. For example, in \eqref{pathint-maxwell-2} if we would have $J_{\mu} =0$ as the only configuration this would give us the free Maxwell theory. Another example is to consider $J_{\mu}$ as a continuous field $(\sum_{J} \rightarrow \int {\cal D} J)$, as a result we have that $J_{\mu}$ turns into a Lagrange multiplier forcing the gauge field to vanish, which is just the Meissner effect in a perfect superconductor (with zero penetration {length}). 
% Consider $J_\mu$ as a continuous field represents the condensation process. 
% The former case it is called  condensation process and it is thus an operation that maps an ensemble of $1$-currents into an ensemble of $1$-forms. The first one is the dilution process. This operation (condensation/dilution) specifies a physical process that connects different theories, describing the system in different phases. 
%\noindent To be more clear we define the sum over currents making explicit the weight $e^{-S_J}$ defining the ensemble of currents. 

\noindent Supplementing \eqref{pathint-maxwell-2} with contact terms to the classical current. The contact terms are  on the action $S_{J}$. The new partition function reads
%Also, we insert in the path integral the unit 
% \begin{align}
% \label{unitpoisson}
% \int {\cal D }\eta\; \delta[J^{\mu}(x) - \eta^{\mu}(x)] = 1
% \end{align}
% obtaining
\begin{widetext}
\begin{align}
\label{genfunc}
  Z[j]  = \sum_{J}  \int {\cal D} A {\cal D} \theta  e^{-\int d^4x \left( \frac{1}{4} F_{\mu\nu}F_{\mu\nu} - iq \left( A_{\mu} + \frac{1}{q} \partial_{\mu}\theta \right) J_{\mu}\right)} e^{-ie\int d^4x \; A_{\mu}j_{\mu}  }e^{-S_{J}} 
\end{align}
\end{widetext}
%where, due to delta function, we replaced $J\rightarrow \eta$ everywhere in the argument. Now,  using the Poisson summation formula (\ref{poissonformula}), we obtain the equivalent formulation
%\begin{align}
%\label{genfunc2}
%Z[j]  = \sum_{K}  \int {\cal D} A {\cal D} \theta {\cal D} \eta e^{-\int d^4x \left( \frac{1}{4} F_{\mu\nu}F_{\mu\nu} - iq \left( A_{\mu} + \frac{1}{q} \partial_{\mu}\theta + \frac{2\pi}{q} K_{\mu}\right) \eta_{\mu}\right)} e^{-ie\int d^4x \; A_{\mu}j_{\mu}  }e^{-S_{\eta}} 
%\end{align}

%\noindent We can now have a more clear understanding of the meaning of the ensemble action $S_{J}$. In order to perform the integral over $J$ we have to specify a form for this action. 
%Consider first the simplest choice of putting $S_{J} = 0$. In this case, $\eta$ is a Lagrange multiplier that imposes 
%\begin{align}
%\label{etalag}
% A_{\mu} = -\frac{1}{q} \partial_{\mu}\theta - \frac{2\pi}{q} K_{\mu} 
%\end{align}
%since $K$ is a $1$-current, it defines localized lines in space. This equation just says that the gauge field $A_{\mu}$ is restricted to flux filaments and thus we see that $K$ stands for vortices in the system. If $K$ dilutes the vortices disappears and the system becomes a perfect superconductor with zero penetration length for the magnetic field. On the other hand, if $K$ condenses then we recover free Maxwell theory with the field $K$ as the gauge field. 
\noindent The allowed local terms for the current looks like:
\begin{align}
 \label{Seta}
  S_{J} = \frac{1}{2m^2} J_{\mu}J^{\mu} + \frac{1}{2m^2 m^2_1} J_{\mu}\Box J^{\mu} + \cdots
\end{align}
This is a derivative expansion where the first term represents the lowest energy fluctuation of $J_\mu$. From \eqref{delta}, {one can express the current} as $J_{\mu}=\epsilon_{\mu\nu\lambda\rho}\partial^{\nu}\Sigma^{\lambda\rho}$.
%, where $\epsilon$ is the Levi-Civita symbol. 
In this sense $J_{\mu}J^{\mu}$ can be {treated} as kinetic term. Considering only the low energy fluctuations in \eqref{Seta} in the condensate phase and integrating \eqref{genfunc} in $J$ we have
%\begin{align}
%\label{eqmoteta}
%J_{\mu} =  iq m^2 \left( A_{\mu} + \frac{1}{q} \partial_{\mu}\theta\right)
%\end{align}
% and we obtain 
\begin{widetext}
 \begin{align}
 \label{dilutecharges3}
 Z[j] = \int {\cal D} A \int {\cal D} \theta\;  e^{-\int d^4x \left( \frac{1}{4} F_{\mu\nu}F^{\mu\nu} - \frac{q^2m^2}{2}\left(  A_{\mu} + \frac{1}{q} \partial_{\mu}\theta\right)^2 \right)} e^{-ie\int d^4x \; A_{\mu}j_{\mu}  } 
\end{align}
\end{widetext}
Thus we have {obtained} the action for the electromagnetic response in a superconductor with penetration length $\sim 1/m$. The condensation of {currents} drove the system to a phase transition. In \cite{Braga:2016atp}, the authors showed that the dilution/condensation process can also be described in a dual point of view. The dual of the charge condensation described here is the dilution of the equivalent defects (vortices) in the superconductor. Meaning that if the vortices dilute ({and} charge condensation occurs), the system becomes a perfect superconductor as in \eqref{dilutecharges3} or if the vortices condensate ({and} charge dilution takes place) we recover the free Maxwell theory. As it is known, the superconductivity phase of a material can be {destroyed} by proliferation {of vortices}, and, as it is shown in \cite{Braga:2016atp}, the picture presented here, describes this scenario. The dual picture of \eqref{dilutecharges3} can be obtained by introducing  
\begin{equation}
\int {\cal D}\eta~\delta[J^\mu(x)-\eta^\mu(x)]=1
\label{delta1}
\end{equation}
in the path integral \eqref{genfunc}, using the Poisson identity
\begin{equation}
\label{poissonid}
\sum_J \delta[J^\mu(x)-\eta^\mu(x)]=\sum_Ke^{i2\pi\int d^4x\eta^\mu K_\mu}
\end{equation}
and by integrating $\eta_\mu$ we obtain
\begin{widetext}
 \begin{align}
 \label{dilutecharges3dual}
 Z[j] =\sum_K \int {\cal D} A \int {\cal D} \theta\;  e^{-\int d^4x \left( \frac{1}{4} F_{\mu\nu}F^{\mu\nu} - \frac{q^2m^2}{2}\left(  A_{\mu} + \frac{1}{q} \partial_{\mu}\theta+\frac{2\pi}{q}K_\mu\right)^2 \right)} e^{-ie\int d^4x \; A_{\mu}j_{\mu}  } 
\end{align}
\end{widetext}
From \eqref{dilutecharges3dual} it can be seen that, if $K$ becomes a continuous field, {the} integration over $K$ drives the system to a free Maxwell theory. The Poisson identity \eqref{poissonid} is the key idea to implement and to understand the dual scenario: from \eqref{poissonid}, as an example, it can be seen that if we have $J_\mu=0$ as the only configuration, this identity will be reduced to $\delta(\eta_\mu)=\int{\cal D}K  e^{i2\pi\int d^4x\eta^\mu K_\mu}$. This means that for a charge dilution ($J_\mu=0$) the result will be that the $K$ condensates $(\sum_K\to\int{\cal D}K)$. If the $J$ proliferates $(\sum_J\to\int{\cal D}J)$ this makes the l.h.s. of \eqref{poissonid} {to go} to $1$, forcing $K=0$ on the {r.h.s.} of the equality. This points that $K_\mu$ are defects which are dual to the charges $J_\mu$. Due to \eqref{delta1}, the dual version of  \eqref{Seta} reads
\begin{align}
 \label{Seta1}
  S_{\eta} = \frac{1}{2m^2}\eta_{\mu}\eta^{\mu} + \frac{1}{2m^2 m^2_1}\eta_{\mu}\Box \eta^{\mu} + \cdots
\end{align}
and in order to obtain \eqref{dilutecharges3dual} we  have {only} considered the lowest energy fluctuation in $\eta$ as in the previous case with $J$. As it was pointed before, if $K$ condensates we have free Maxwell theory. In the next section we will show that {for the case of} high energy fluctuations {of} $\eta$, even when $K$ condensates, local Maxwell theory will not be recovered anymore.
%%==============================================================================
%\section{The Gribov-like term}
%%==============================================================================
%Unlike in \eqref{Seta} where a derivative expansion were considered here in this section we want to chose only allowed local terms in $S(J)$. We chose only
%\begin{align}
%  S_{J} = \frac{1}{2m^4} J_{\mu}\partial^2 J^{\mu}
%\end{align}

%==============================================================================
\section{Podolsky electrodynamics as an emergent theory}
\label{emergentnonlocalterm}
%==============================================================================
From the previous section, it can be seen that by considering  low energy fluctuations of $\eta_\mu$ we are able to describe the electromagnetic response in a superconductor. In this section, our discussion begins around the equation \eqref{Seta1} and explore what is the gauge emergent system by considering high energy fluctuations of $\eta_\mu$. Consider high energy fluctuations of  $\eta_\mu$ means that 
%From \eqref{dilutecharges3dual}, we know in advance that new system will emerge only in the condensate phase. 
\begin{eqnarray}
 \label{Seta2}
 S_{\eta}&=&\frac{1}{2m^2} \eta_{\mu}\eta^{\mu} + \frac{1}{2m^2 m^2_p} \eta_{\mu}\Box \eta^{\mu} + \cdots\nonumber\\
&=&  \frac{1}{2m^2} \eta_{\mu}\left[\sum_{n=0}^\infty\left(\frac{\Box}{m^{2}_p}\right)^n\right]\eta_{\mu}\nonumber\\
&=&  \frac{1}{2m^2} \eta_{\mu}{\cal F}\left({\Box}/{m^{2}_p}\right)\eta_{\mu}
\end{eqnarray}
where $m_p$ stands for the Podolsky mass as in \cite{Bonin:2019xss}. Then the integration in $\eta_\mu$ generates:
\begin{widetext}
 \begin{align}
 Z[j] =\sum_K \int {\cal D} A \int {\cal D} \theta\;  e^{-\int d^4x \left( \frac{1}{4} F_{\mu\nu}F^{\mu\nu} - \frac{q^2m^2}{2{\cal F}\left({\Box}/{m^{2}_p}\right)}\left(A_{\mu} + \frac{1}{q} \partial_{\mu}\theta+\frac{2\pi}{q}K_\mu\right)^2 \right)} e^{-ie\int d^4x \; A_{\mu}j_{\mu}  } 
 \end{align}
\end{widetext}
At this point, we will redefine the gauge field for $A'_\mu=A_{\mu} + \frac{1}{q} \partial_{\mu}\theta+\frac{2\pi}{q}K_\mu$. This redefinition maintains the electromagnetic field strength tensor expression preserved. The action then reads
%In the Lorentz gauge the action reads:
\begin{eqnarray}
\label{themodel}
S&=&\int d^4x \left(\frac{1}{4} F_{\mu\nu}F^{\mu\nu} - A'_\mu\frac{q^2m^2}{2{\cal F}\left({\Box}/{m^{2}_p}\right)}A'_\mu \right)%+\frac{1}{2\alpha}\left(\partial_\mu A_\mu\right)^2\right)\nonumber\\
%&=&\int d^4x \left(\frac{1}{2} A_\mu\left(-\left(\partial^2+\frac{m^2}{F\left({\partial^2}/{m^{2}}\right)}\right)\delta_{\mu\nu}+\left(1-\frac{1}{\alpha}\right)\partial_\mu\partial_\nu\right)A_\nu \right)
\end{eqnarray}

 It is interesting to note the following effect.
Under an appropriate change of variables, the nonlocality in the kinetic term can be transferred to the vertices (or mass term), and vice versa. It can be done in the following manner.

Let us consider the Lagrangian
\begin{eqnarray}
{\cal L}&=&\frac{1}{2}\pa_m\phi e^{\frac{\Box}{\mu^2}}\pa^m\phi-V(\phi).
\end{eqnarray}
Here, the nonlocality is concentrated in a kinetic term.
Let us do the change of variables
\begin{eqnarray}
e^{\frac{\Box}{2\mu^2}}\phi\to\tilde{\phi}.
\end{eqnarray}
Our Lagrangian takes the form
\begin{eqnarray}
{\cal L}&=&\frac{1}{2}\pa_m\tilde{\phi}\pa^m\tilde{\phi}-V(e^{-\frac{\Box}{2\mu^2}}\tilde{\phi}).
\end{eqnarray}
So, the potential becomes nonlocal instead of the kinetic term (in certain cases when the kinetic term looks like $\phi\Box\hat{T}\phi$, and the potential term looks like $((\hat{T})^{1/2}\phi)^n$, with $\hat{T}$ is an operator introducing the nonlocality, f.e. $\hat{T}=e^{\frac{\Box}{2\mu^2}}$ as in the example above, we can remove the nonlocality both from kinetic and potential term, but these cases are trivial).

The similar situation takes place for other field theory models, including the electromagnetic field. The model (\ref{themodel}), under the replacement ${\cal F}\left({\Box}/{m^{2}}\right)^{-1/2}A'_\mu\to\tilde{A}_{\mu}$ becomes
\begin{eqnarray}
\label{themodel1}
S=\int d^4x \left(\frac{1}{4} F_{\mu\nu}{\cal F}\left({\Box}/{m^{2}_p}\right)F^{\mu\nu} - \frac{q^2m^2}{2}A'_\mu A'_\mu \right).%+\frac{1}{2\alpha}\left(\partial_\mu 
\end{eqnarray}
%So, we generated the essentially nonlocal Maxwell term.
We note that within all these our replacements by the rule $\phi\Box\hat{L}\phi$, with $\hat{L}$ is the nonlocal operator, and the same rule for $A_{\mu}$, no extra contributions to the effective actions are generated. Indeed, when we carry out these transformations,  although there are nonlocal, they are linear in fields, so, in the generating functional we get only the extra multiplier $\det\hat{L}^{1/2}$, and since it does not depend on any fields, it yields only a field-independent additive term in the effective action which clearly can be neglected. 
So, we generated the essentially non-local Podolsky electrodynamics with the action
\begin{widetext}
 \begin{align}
\label{themodel1a}
 Z[j] =\sum_K \int {\cal D} A \int {\cal D} \theta\;  e^{-\int d^4x \left(\frac{1}{4} F_{\mu\nu}F^{\mu\nu}+\frac{1}{4m_p^2}\partial^\mu F_{\mu\nu}\left[\sum\limits_{n=0}^\infty\left(\frac{\Box}{m^{2}_p}\right)^n\right]\partial_\beta F^{\beta\nu}- \frac{q^2m^2}{2}A'_\mu A'_\mu \right)}\cdot e^{-ie\int d^4x \; A_{\mu}j_{\mu}  }
 \end{align}
\end{widetext}
%\begin{widetext}
%\begin{eqnarray}
%\label{themodel1a}
%S&=&\int d^4x \left(\frac{1}{4} F_{\mu\nu}F^{\mu\nu}+\frac{1}{4} F_{\mu\nu}\left[\sum_{n=1}^\infty\left(\frac{\partial^2}{m^{2}_p}\right)^n\right]F^{\mu\nu}- \frac{q^2m^2}{2}A'_\mu A'_\mu \right)\nonumber\\
%&=&\int d^4x \left(\frac{1}{4} F_{\mu\nu}F^{\mu\nu}+\frac{1}{4m_p^2}\partial^\mu F_{\mu\nu}\left[\sum_{n=0}^\infty\left(\frac{\partial^2}{m^{2}_p}\right)^n\right]\partial_\beta F^{\beta\nu}- \frac{q^2m^2}{2}A'_\mu A'_\mu \right).
%\end{eqnarray}
%\end{widetext}
 where an integration by parts was used to obtain the result. It can be seen from \eqref{themodel1a} that even if $K$ condensates, {which corresponds to} integration over $K$, the Podolsky term will not vanish. It is worth mention that if we have considered only the first two terms in the expansion \eqref{Seta2}, the resulting action would be the local Podolsky action, which it would be equivalent to only consider the term $n=0$ is the sum in \eqref{themodel1a}. Therefore we show that the  Podolsky electrodynamics can be obtained using a mechanism that it was originally used to describe phase transitions due to the condensation/proliferation of topological defects.

It is interesting to note that the higher-derivative Podolsky term can be generated as well within the usual perturbative approach. Let us start with the {usual} Lagrangian of the spinor field coupled to the electromagnetic one:
\be
{\cal L}=\bar{\psi}(i\ds-e\As-m)\psi.
\ee
The one-loop effective action of the gauge field is immediately written in the form of the fermionic determinant:
\be
\Gamma^{(1)}=i{\rm Tr}\ln(i\ds-e\As-m).
\ee
We consider the simplest contribution to it generated by the two-point function of $A^{\mu}$:
\be
\Gamma^{(1)}_2=-\frac{e^2}{2}\int d^4p A_{\mu}(-p)\Pi^{\mu\nu}(p)A_{\nu}(p),
\ee
where
\be
\label{trace}
\Pi^{\mu\nu}(p)={\rm tr}\int\frac{d^4k}{(2\pi)^4}\gamma^{\mu}\frac{1}{\ks-m}\gamma^{\nu}\frac{1}{\ks+\ps-m}.
\ee
It is well known that the effective action itself is nonlocal being an infinite series in derivatives of the external fields, or as is the same, in the external momentum $p$. While the contribution of the second order in an external momentum became a paradigmatic result in QED describing the wave function renormalization of the gauge field, the higher-order results have not been discussed up to now. So, we expand (\ref{trace}) up to the fourth order and find
\ben
\Pi_4^{\mu\nu}(p)&=&{\rm tr}\int\frac{d^4k}{(2\pi)^4}\gamma^{\mu}\frac{1}{\ks-m}\gamma^{\nu}\frac{1}{\ks-m}\ps\frac{1}{\ks-m}\times\nonumber \\&\times& \ps\frac{1}{\ks-m}\ps\frac{1}{\ks-m}\ps\frac{1}{\ks-m}.
\een
\vspace*{1mm}

The trace and integral can be calculated explicitly. We arrive at
\ben
\Pi_4^{\mu\nu}(p)&=&\frac{4p^2}{15m^2(4\pi)^2}(p^{\mu} p^{\nu}-p^2\eta^{\mu\nu}).
\een
The corresponding contribution to the effective action is
\ben
\Gamma^{(1)}_{2,4}=-\frac{e^2}{15m^2(4\pi)^2}F_{\mu\nu}\Box F^{\mu\nu}.
\een
This term is finite as it must be. It has just the desired Podolsky form. In principle, the contributions to the effective action involving sixth and higher even orders in momenta can be obtained as well, so, one can write down the complete one-loop two-point function as
\ben
\Gamma^{(1)}_2=e^2F_{\mu\nu}\left(\sum\limits_{n=0}^{\infty}c_n(\frac{\Box}{m^2})^n\right) F^{\mu\nu},
\een
where the zero order is the known renormalized QED result. In principle, the function 
$$
f(\Box)=\sum\limits_{n=0}^{\infty}c_n\left(\frac{\Box}{m^2}\right)^n
$$
can be defined, where $c_n$ are some numbers, however, apparently {this function} can be found only order by order but not in the closed form. 

\section{Higher-derivative Chern-Simons term}\label{sec-cs}

In this section, we show that by extending the mechanism presented in the section \ref{secI}, for a Chern-Simons (CS) theory, and following the prescription in the section \ref{emergentnonlocalterm}, the corresponding emergent theory,  obtained along the same lines as Eq. \eqref{themodel1},  is a {higher-derivative} CS theory. For the CS theory the analog of \eqref{maxwell} is 
\begin{equation}
\label{csaction}
S_{cs} = \int d^3x \left( \frac{\kappa}{2} \epsilon_{\mu\nu\rho} A_\mu\partial_\nu A_\rho - iqA_{\mu}J_{\mu}  \right),
\end{equation}
and, {since} the CS action is gauge invariant, \eqref{delta} still holds. Following the prescriptions presented in the previous sections we arrive at
\begin{eqnarray}
S_{cs} &=& \int d^3x \Big( \frac{\kappa}{2} \epsilon_{\mu\nu\rho} A_\mu\partial_\nu{\cal F}\left({\Box}/{m^{2}_{cs}}\right) A_\rho-\nonumber\\ &-&
\frac{q^2m^2}{2}A'_\mu A'_\mu   \Big).
\end{eqnarray}
This is the analog of \eqref{themodel1} for the CS theory.

\section{Comments and conclusions}\label{sec-com}

We have succeeded to generalize the Julia-Toulouse mechanism {to} a case of {a} nonlocal contact term. One should note that {\it a priori}, there are no restrictions on the form of the contact term within this approach. We demonstrated explicitly that in this case, one can generate a nonlocal generalization of the electrodynamics whose action is an infinite series in derivatives, so that one has, besides of the usual Maxwell term, also the Podolsky term, {and higher-order terms}.  We showed explicitly that the Podolsky term can be generated as a quantum correction as well being finite, so, we can see that perturbative and non-perturbative approaches for obtaining new terms are equivalent in a certain sense as it has been claimed in \cite{CWours}. Actually, we demonstrated that the Julia-Toulouse approach opens {broad} possibilities for obtaining new effective theories with higher-derivative operators. It could be interesting to generalize this approach for Lorentz-breaking case since earlier, the Julia-Toulouse methodology has been successfully applied in the Lorentz-breaking case in {the} three-dimensional theory \cite{CWours}. Especially, it is interesting to study the impacts of the dimension-six terms considered in \cite{Ferr}, within the Julia-Toulouse approach. Besides the possible applications within Lorentz symmetry breaking scenarios, this approach can also be used to show how magnetic permeability arises from a condensation of topological defects. We expect to do these studies in forthcoming papers. 

\vspace*{1mm}

\section*{Acknowledgments}
The authors thank CNPq  for financial support. The work by A.Yu.P. has been supported by CNPq project 303783/2015-0.  P. J. Porf\'{i}rio would like to thank the Brazilian agency CAPES for financial support (PDE/CAPES grant, process 88881.17175/2018-01) and Department of Physics and Astronomy, University of Pennsylvania, for the hospitality.
%%%%%%%%%%%%%%%%%%%%%%%%%%%%%%%%%

\end{document}